\documentclass[prb,twocolumn,superscriptaddress,showpacs,amssymb]{revtex4}

\usepackage{psfrag,graphicx}
\usepackage{dcolumn}
\usepackage{bm}
\usepackage{amsfonts,amssymb,amsmath} 
\usepackage{xcolor}
\usepackage{soul}
\usepackage{epstopdf}
\usepackage{mathbbol}

\newcommand{\be}{\begin{equation}}
\newcommand{\ee}{\end{equation}}
\newcommand{\bq}{\begin{eqnarray}}
\newcommand{\eq}{\end{eqnarray}}

\newcommand{\rf}[1]{(\ref{#1})}

\newcommand{\tr}{\mathrm{tr}}

\newcommand{\en}{\varepsilon}
\newcommand{\baleqn}{\begin{equation}\begin{aligned}}
\newcommand{\ealeqn}{\end{aligned}\end{equation}}
\newcommand{\baleqns}{\begin{equation*}\begin{aligned}}
\newcommand{\ealeqns}{\end{aligned}\end{equation*}}

\begin{document}
\title{Conformal energy currents on the edge of a topological superconductor}

\author{Chris N. Self}
\email{cns08@ic.ac.uk}
\affiliation{School of Physics and Astronomy, University of Leeds, Leeds, LS2 9JT, United Kingdom}
\affiliation{Department of Physics, Imperial College London, London, SW7 2AZ, United Kingdom}
\author{Jiannis K. Pachos}
\affiliation{School of Physics and Astronomy, University of Leeds, Leeds, LS2 9JT, United Kingdom}
\author{James R. Wootton}
\affiliation{Department of Physics, University of Basel, Klingelbergstrasse 82, CH-4056 Basel, Switzerland}
\author{Sofyan Iblisdir}
\affiliation{Departamento de An\'alisis Matem\'atico, Facultad de Matem\'aticas, Universidad Complutense de Madrid, 28040 Madrid, Spain}


\date{\today}
\pacs{74.25.F-, 03.65.Vf, 11.25.Hf}

\begin{abstract}
The boundary of a 2D topological superconductor can be modeled by a conformal field theory. Here we demonstrate the behaviors of this high level description emerging from a microscopic model at finite temperatures. To achieve that, we analyze the low energy sector of Kitaev's honeycomb lattice model and probe its energy current. We observe that the scaling of the energy current with temperature reveals the central charge of the conformal field theory, which is in agreement with the Chern number of the bulk. Importantly, these currents can discriminate between distinct topological phases at finite temperatures. We assess the resilience of this measurement of the central charge under coupling disorder, bulk dimerisation and defects at the boundary, thus establishing it as a favorable means of experimentally probing topological superconductors.

\end{abstract}

\maketitle

\section{Introduction}
Topological superconductivity is a uniquely secretive phase of matter.
Such materials do not conduct charge currents, expel magnetic fields and their topological signatures are hidden from any local observable \cite{TSC}. 
Through an effective gravitational description it has been shown that this ``dark matter" of solid-state physics gives rise to a conformal field theory (CFT) at its boundary~\cite{Gian}. 
This CFT description of the topological edge states is remarkably robust. 
Unlike genuinely $1+1$-dimensional CFT, associated with fine-tuned critical points, these edge CFT typically persist across finite regions of the superconductor parameter space. 
This makes such systems an exciting medium for investigating direct signatures of conformal invariance both theoretically and experimentally.

There is a tight relation between the Chern number $\nu$, describing the bulk physics, and the central charge $c$ of the edge CFT~\cite{Gian}, namely $c=\nu/2$. 
At small temperatures, $T$, conformal field theory predicts an energy current mediated by the topological edge states~\cite{cappelli,Furusaki2013,Sumiyoshi2012}, which scales as $I_\text{CFT} = \frac{\pi}{12}\,c\,T^2$. 
Practically however, the edge states are not perfectly isolated from the rest of the system. 
They have a finite penetration into the bulk, which has its own thermal behavior. 
Inevitably these behaviors will mix. 
A natural question is whether it is still possible to obtain conclusive signatures of CFT thermal properties.

These currents then have the potential to characterize topological phases that remain robust at finite temperature.
Identifying topological phases at finite temperature has recieved attention in its own right through the definition of the topological Uhlmann number~\cite{viyuela2015}, which generalises the notion of a Chern number computed from the band structure to finite temperature.
In the Quantum Hall setting, approaches have also been developed to compute topological conductivites at finite temperature and in the presence of disorder~\cite{song2014,xue2012}.

Here we investigate the edge physics from a microscopic description of a topological superconductor. 
As a concrete example we study Kitaev's honeycomb model \cite{Kitaev06}: a 2D spin liquid that supports topological superconducting phases with a variety of Chern numbers~\cite{Ville,VilleVortex, Lahtinen-Pachos-Chern}. 
Due to the analytical tractability of this model, it is amenable to a wide variety of numerical studies such as finite temperature analysis~\cite{Nasu2015,Nasu2015b,Metavitsiadis2016}. 
We demonstrate that the energy currents, $I(T)$, can be given in terms of two-point fermionic correlators and we investigate their behavior for various phases of the honeycomb lattice model. 
We identify the range of temperatures for which the currents obeys the CFT prediction and show how to identify the central charge $c$ of the CFT. 
We see that $I(T)$ can be used to cleanly signal a finite temperature topological phase transition, when the system parameters vary. 
In addition, we explicitly show the topological nature of the central charge by studying the resilience of $I(T)$ to random disorder and boundary defects. This establishes the energy current as the natural observable for theoretically and experimentally probing topological superconductors at finite temperature.


\section{Energy currents in Kitaev's honeycomb model} 
The Kitaev honeycomb model is defined for spin-1/2 particles at the vertices of a honeycomb lattice~\cite{Kitaev06}. 
The spins interact with their nearest-neighbors anisotropically, with couplings $J_x$, $J_y$ and $J_z$, where we set $J_x = J_y = J$  and $J_z = 1$. 
A weak three-body interaction with coupling $K$, representing the effects of a magnetic field perturbation, breaks time-reversal symmetry. 
{The total Hamiltonian of the model is}
\baleqn
H &= -\,J\sum_{x-\textrm{links}} \sigma_x^i \sigma_x^j -J\sum_{y-\textrm{links}} \sigma_y^i \sigma_y^j \\ 
&\quad -\sum_{z-\textrm{links}} \sigma_z^i \sigma_z^j -K\sum_{(i,j,k)} \sigma_x^i \sigma_y^j \sigma_z^k \, ,
\label{eqn:spinHam}
\ealeqn
{where the three-body subsets $(i,j,k)$ are made up of sets of three adjacent spins and commute with the two-body terms.}
The model supports two types of gapped excitations: fermions and vortices. 
In the basis of vortex excitations the Hamiltonian is block-diagonal. 
Each block, corresponding to a particular vortex configuration, is called a \emph{vortex sector}, $\mathnormal{V}$. 
{The remaining degrees of freedom within each sector can be expressed in terms of a free Majorana-fermion} ($c_i^\dagger = c_i$, $c_i^2 = 1$) {Hamiltonian,}
\begin{equation}
H_\mathnormal{V} = \frac{i}{4} \sum_{i,j} {A}_{ij} c_i c_j  \, 
\label{eqn:majoranahamiltonian}
\end{equation}
{where $A_{ij}$ is a real antisymmetric matrix encoding the vortex data and is a function of $J$ and $K$.}
For particular choices of $J$, $K$ and $\mathnormal{V}$ the free fermions support topological phases with different Chern numbers\cite{Kitaev06,Ville}. 
Here we focus on the no-vortex (NV) sector with $\nu_\text{NV}=1$, the full-vortex (FV) sector with $\nu_\text{FV} = 2$, as well as a toric code (TC) phase with $\nu_\text{TC} = 0$.

\begin{figure}[t]
\centering
\includegraphics[width=0.9\columnwidth]{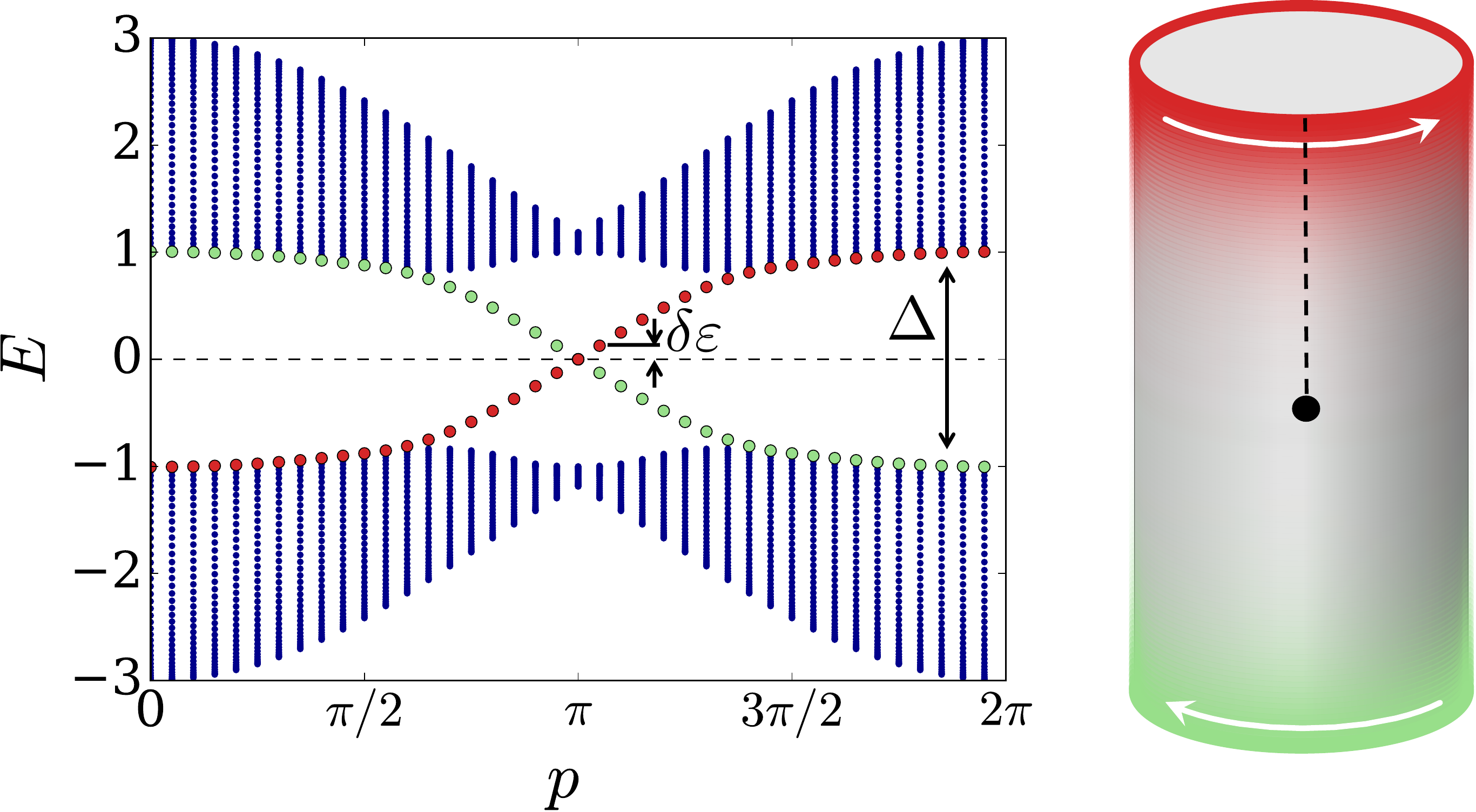}
\caption{Dispersion relation of the edge modes (Left) that are exponentially localized on opposite ends of the cylinder (Right) moving in opposite directions. These modes are witnessed as two midgap bands that cross at $p = \pi$. The energy gap, $\Delta$, of the model and the discretization gap, $\delta \varepsilon$, due to the finite size $L$ of the system are shown.}
\label{fig:cyl}
\end{figure}

Consider the honeycomb lattice in a phase with non-trivial Chern number, wrapped around a cylinder with a perimeter of length $L$ and height $D$. 
The non-zero Chern number dictates that the spectrum contains chiral midgap modes. 
These modes have linear dispersion and are exponentially localized on the boundaries of the system, as shown in Fig.~\ref{fig:cyl} where the color intensity is drawn decaying exponentially into the bulk. 
Thermal excitation of these modes give rise to an energy current on the edge. 
It is possible to heuristically estimate this current. At low temperatures $T$ compared to the bulk energy gap $\Delta$, it is given by 
\be
\label{eqn:current2}
I_\text{edge} \approx \sum_{p:\,\varepsilon(p) \geq \delta\varepsilon }^{\varepsilon(p) \leq \Delta} n_\beta(\varepsilon) \, \varepsilon(p) \, \frac{\delta \varepsilon}{2 \pi}  \,, 
\ee
where $n_\beta(\varepsilon)=1/(1 + e^{\beta \varepsilon})$ gives the fermionic occupancies and $\varepsilon(p)$ is the edge state dispersion relation, shown in Fig.~\ref{fig:cyl}~(Left). 
In the limit of infinite size the edge modes of the cylinder are gapless around $p = \pi$. A finite circumference $L$ induces an infra-red cutoff given by $\delta \varepsilon = O(1/L)$. 
Additionally, $\Delta$ naturally sets an ultra-violet cutoff \footnote{{$\Delta$ is evaluated by imposing periodic boundary conditions in both directions.}}. 
These conditions determine the limits of the sum Eqn.~\eqref{eqn:current2}. If we send $\delta \varepsilon \rightarrow 0$ and $\Delta \rightarrow \infty$ then Eqn.~\eqref{eqn:current2} evaluates to the CFT current, $I_\text{CFT}$, exactly. 
However, these limits place bounds on the range of temperatures at which we expect to see CFT current behavior, given by
\be
O(L^{-1})\ll T\ll\Delta.
\label{eqn:Tcondition}
\ee
Furthermore, Eqn.~\eqref{eqn:current2} assumes perfect distinguishability of the edge modes. 
While this is energetically possible, it cannot be achieved by local position measurements of the current. 
Indeed, while edge modes are exponentially localized they still have a finite penetration into the bulk of the system, as shown in Fig.~\ref{fig:cyl} (Right). 
Hence any attempt to probe them theoretically or experimentally from a microscopic model needs to consider current contributions from all states, not just the midgap modes \cite{Kitaev06}. 
Below we carry out such an analysis and compute the energy currents directly from bulk microscopics. 
We demonstrate that it is still possible to identify CFT behavior from total energy currents. We also identify cases in which the presence of the bulk does significantly change the behavior. 
By studying these currents, the effects of conformal invariance can be directly measured as a response to thermal excitations of the model. 


\begin{figure}[t]
\centering
\includegraphics[width=\columnwidth]{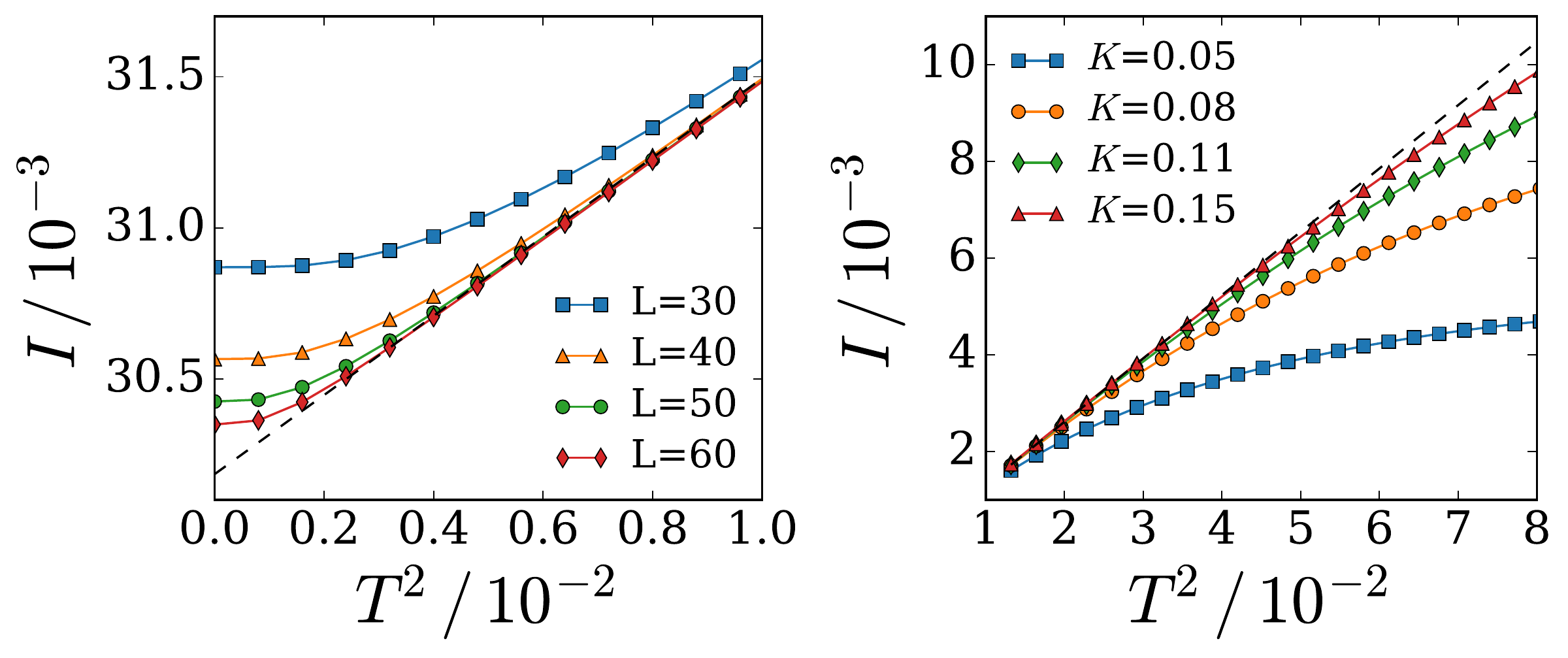}
\caption{Low and high $T$ behavior of $I(T)$.
(Left) In the low-$T$ case increasing system size, $L$, recovers the $T^2$ behavior at lower temperatures. 
(Right) In the high-$T$ case increasing $K \propto \Delta$ delays the divergence of the curves. 
The (Left) panel includes the $I_0$ offset, whereas we have removed it from the (Right) panel. 
In both plots $J=1$. 
In the (Left) plot $K = 0.15$ and in the (Right) plot $L=60$.}
\label{fig:edge-currents-Tbreakdown}
\end{figure}

Let us now define energy currents from a microscopic description. 
For a fixed vortex sector, the spin system reduces to one of free Majorana fermions~\cite{Kitaev06} whose Hamiltonian $H_\mathnormal{V}$ is given in Eqn.~\eqref{eqn:majoranahamiltonian}. 
We separate the Hamiltonian into a sum of terms $h_j$ with support localized about site $j$
\begin{equation}
H_\mathnormal{V} = \sum_j h_j \,\,\,\, \text{with}\,\,\,\, h_j = \frac{i}{4} \sum_{i=1}^N {A}_{ij} c_i c_j  \, .
\label{eqn:spatmajoranahamiltonian}
\end{equation}§
From the Heisenberg equation for dynamics of the $h_j$'s, $\frac{\textrm{d}}{\textrm{d} t} h_j = -i [ H_\mathnormal{V} ,h_j] = -i\sum_k \, [h_k,h_j]$, we can define the current operator $I_{jk}$ as
\be
I_{jk} \equiv -i [h_k,h_j]. 
\label{eqn-current-op}
\ee
To calculate the edge current, we first compute the net energy flux around the cylinder as a function of height $y$. Then we sum up all these local currents between the middle of the system and the boundary, as shown with the dashed line on Fig.~\ref{fig:cyl} (Right), to obtain
\be
I(T) = \sum_{y = D/2}^{D} \, \left(  \sum_{\langle j,k \rangle \,:\, y} \tr( \, \rho_\beta I_{jk} \, ) \right) .
\ee
The inner sum is performed over links $\langle j,k \rangle$ that cross the cut at height $y$, while the outer sum captures the current on one edge only. 
The finite temperature expectation values $\tr( \, \rho_\beta I_{jk} \, )$ are computed from the thermal state $\rho_\beta = {e^{-H_\mathnormal{V}\beta}/ \tr(e^{-H_\mathnormal{V}\beta})}$ ($\beta \equiv 1/T$), which can be obtained by numerically diagonalizing $H_\mathnormal{V}$.

\begin{figure}[t]
\centering
\includegraphics[width=0.75\columnwidth]{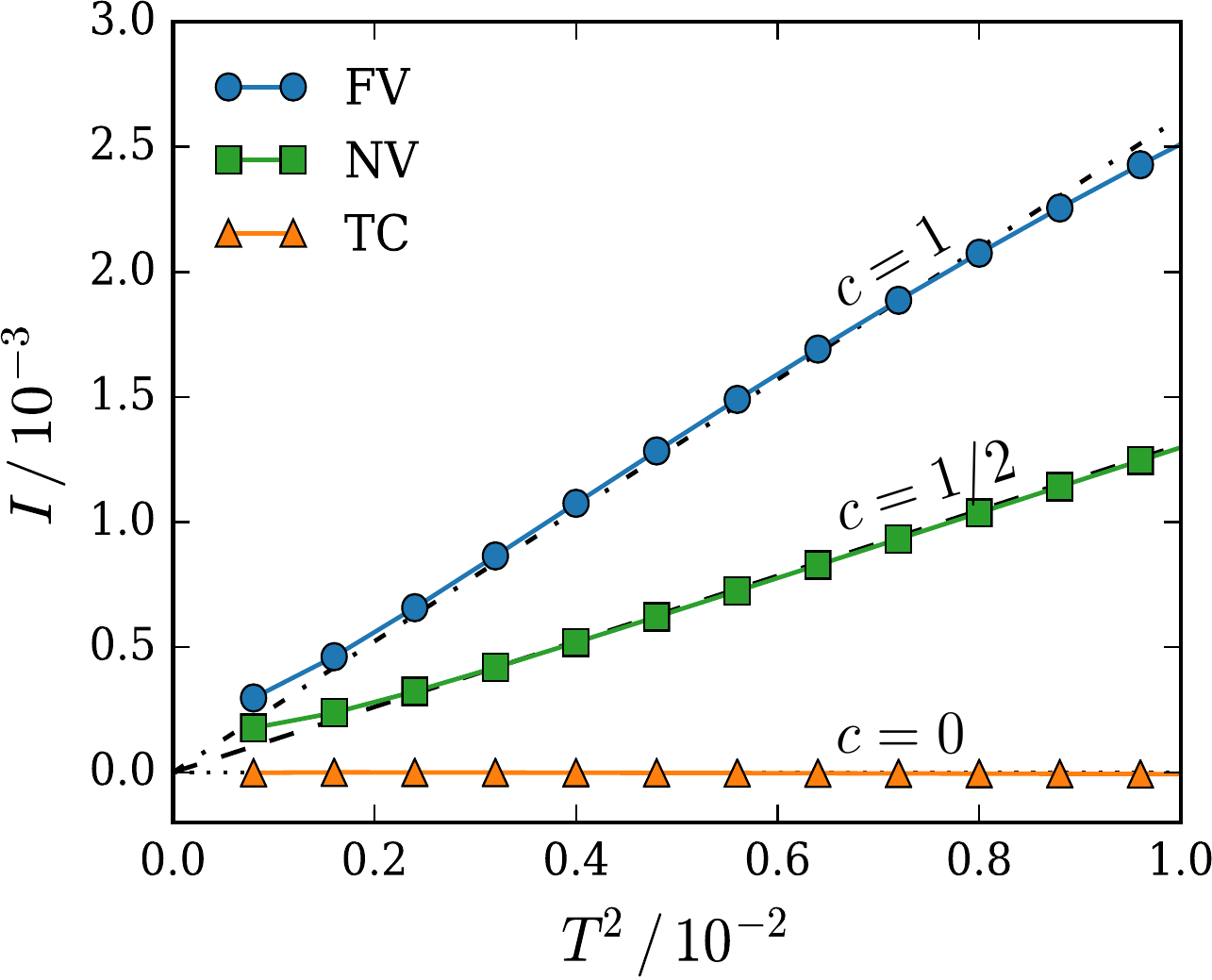}
\caption{Edge current for the no-vortex sector (NV), full-vortex sector (FV) and Toric Code phases (TC) as a function of $T^2$. The currents are shifted to pass through the origin, clearly revealing the $T^2$ scaling. The central charges correspond to dotted line for $c=0$, dashed line for $c=1/2$ and dashed-doted line for $c=1$. In these plots we set $K = 0.15$. For the NV and FV plots we set all $J = 1$, and for the TC plot we take $J = 0.1$. Each is plotted for system size $L = 60$.}
\label{fig:edge-currents}
\end{figure}

\section{Central charge and its topological resilience}
We are now in a position to study the behavior of the energy currents $I(T)$ as we vary the temperature $T$. 
Before we can compare the scaling of $I(T)$ with the CFT prediction, $I_\text{CFT}$, there are two aspects of the energy current we need to address.
These are the contribution from the bulk and the bounds set by condition~\rf{eqn:Tcondition}. 
The prediction of CFT is that the currents should vanish at $T=0$ in the thermodynamic limit, $L \rightarrow \infty$.
By carrying out a finite size scaling analysis for $I(T=0)$ we find that in fact this is not the case due to bulk contributions to the energy current. 

We identify a non-zero value $I_0 = I(T=0,L \rightarrow \infty)$ at zero-temperature, as shown in Fig.~\ref{fig:edge-currents-Tbreakdown} (Left), that depends on the microscopic parameters of the model.
In order to compare the energy currents of different phases we subtract the $I_0$ contribution. 
When condition~\eqref{eqn:Tcondition} is violated we expect $I(T)$ to deviate from the behavior predicted by CFT.
In Fig.~\ref{fig:edge-currents-Tbreakdown} we plot the currents at low-$T$ and high-$T$, for the no-vortex sector.
We see that, in addition to $I_0$, finite size effects shift the value of $I$ at $T=0$. 
When $L$ is increased the currents at $T=0$ tend to $I_0$.
This generic low-$T$ behaviour is seen for all model parameters.
In contrast at large $T$ the currents are sensitive to the bulk gap $\Delta$ but independent of system size. 
We observe that the temperature at which $I$ starts to clearly deviate from $T^2$ scaling (greater than a 10\% difference) grows linearly with $\Delta$.
To summarise, we have verified that the temperature range of interest to find CFT like currents has lower and upper bounds that scale proportional to the inverse system size and the energy gap respectively. 

\begin{figure}[t]
\centering
\includegraphics[width=\columnwidth]{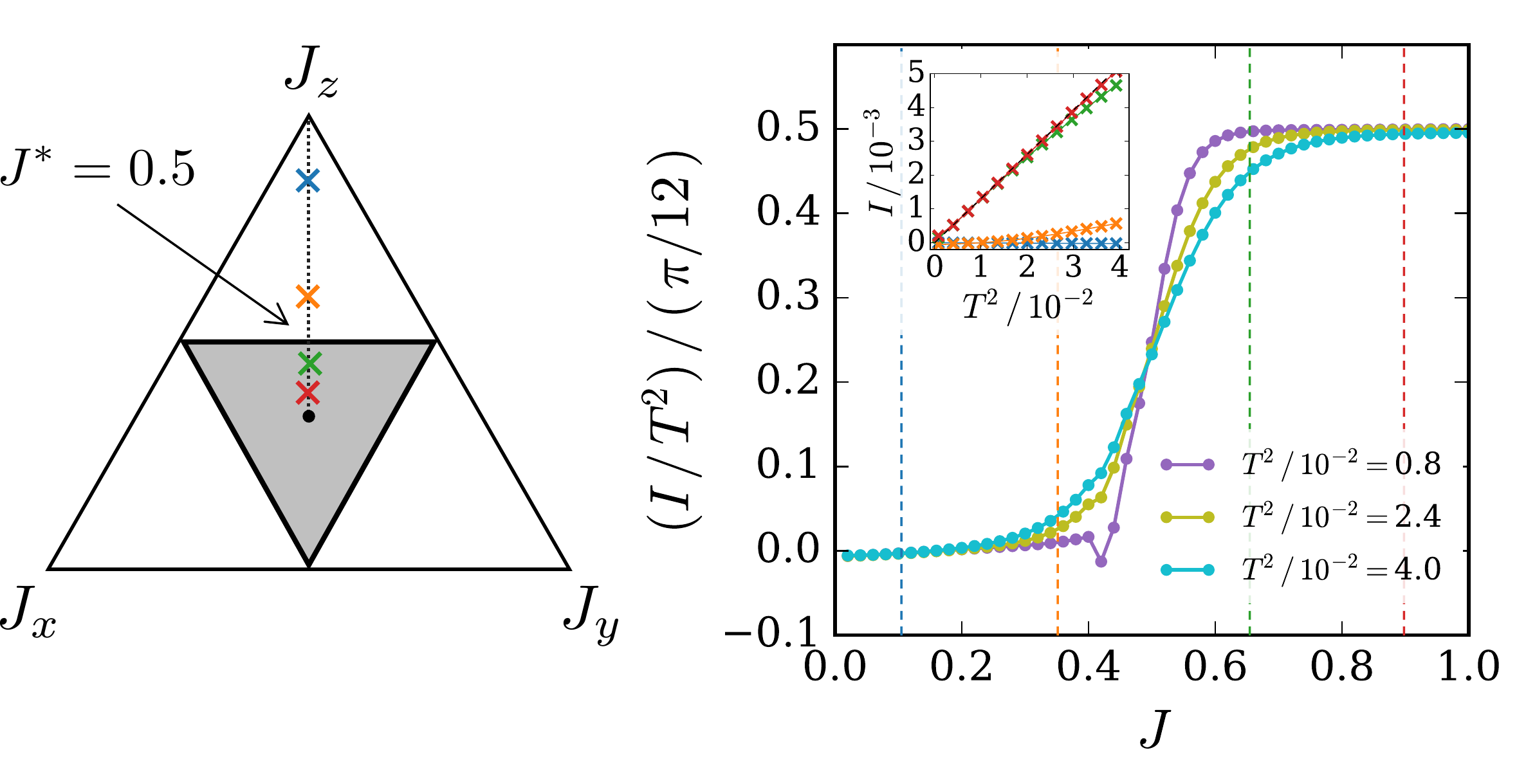}
\caption{
Edge currents indicate a transition between different topological phases at finite temperature. Here we vary $J_x = J_y = J$ for fixed $J_z = 1$, corresponding to the path through parameter space shown on the triangle, {the critical value of $J$ is then $J^* = 0.5$}. Scaled edge currents $I\,/\,T^2$ are plotted against $J$ for different values of $T$. We see a jump through the phase transition that sharpens with decreasing $T$. (Inset) $I$ against $T^2$ plotted for the values $J = 0.1, \, 0.35, \, 0.65, \, 0.9$ indicated on the triangle. The currents are seen to be robust in the topological phase (red and green crosses) and vanish in the Toric code phase (orange and blue crosses). These values of $J$ are indicated on the main plot with vertical lines of the corresponding colors. These plots are for $L=52$ and $K=0.15$.
\label{fig:transition-TC-to_NV}}
\end{figure}

For temperatures that satisfy condition~\eqref{eqn:Tcondition} we find energy currents that behave as CFT currents.
To demonstrate this we plot $I(T)$ against $T^2$ for the Toric code, no-vortex and full-vortex phases in Fig.~\ref{fig:edge-currents}.
We find the $I(T)$s are in excellent agreement with the CFT predictions where for TC, $c=0$, for NV, $c=1/2$ and for FV, $c=1$. 
In this way, we can identify the central charge of the edge theory directly from energy currents.

Since the central charge only takes rational values~\cite{DiF} the current $I(T)$ should jump as we move between two phases with different Chern numbers. 
This gives $I(T)$ the useful theoretical property that it identifies topological phase transitions at finite temperatures. 
In Fig.~\ref{fig:transition-TC-to_NV} (Right) we plot the energy currents as we transition from the TC to NV phases at different temperatures. 
The specific transition we probe is achieved by tuning the $J$ couplings, and is illustrated by the parameter-space diagram on the left of Fig.~\ref{fig:transition-TC-to_NV}. 
Higher temperatures are seen to smear out the transition and at the critical point we find a crossing between the different temperature curves. 
As the finite temperature behavior of the energy currents is uniquely determined by the CFT at $T=0$ the currents are a definitive tool to characterize topological phases at finite temperature.

\begin{figure}[t]
\centering
\includegraphics[width=\columnwidth]{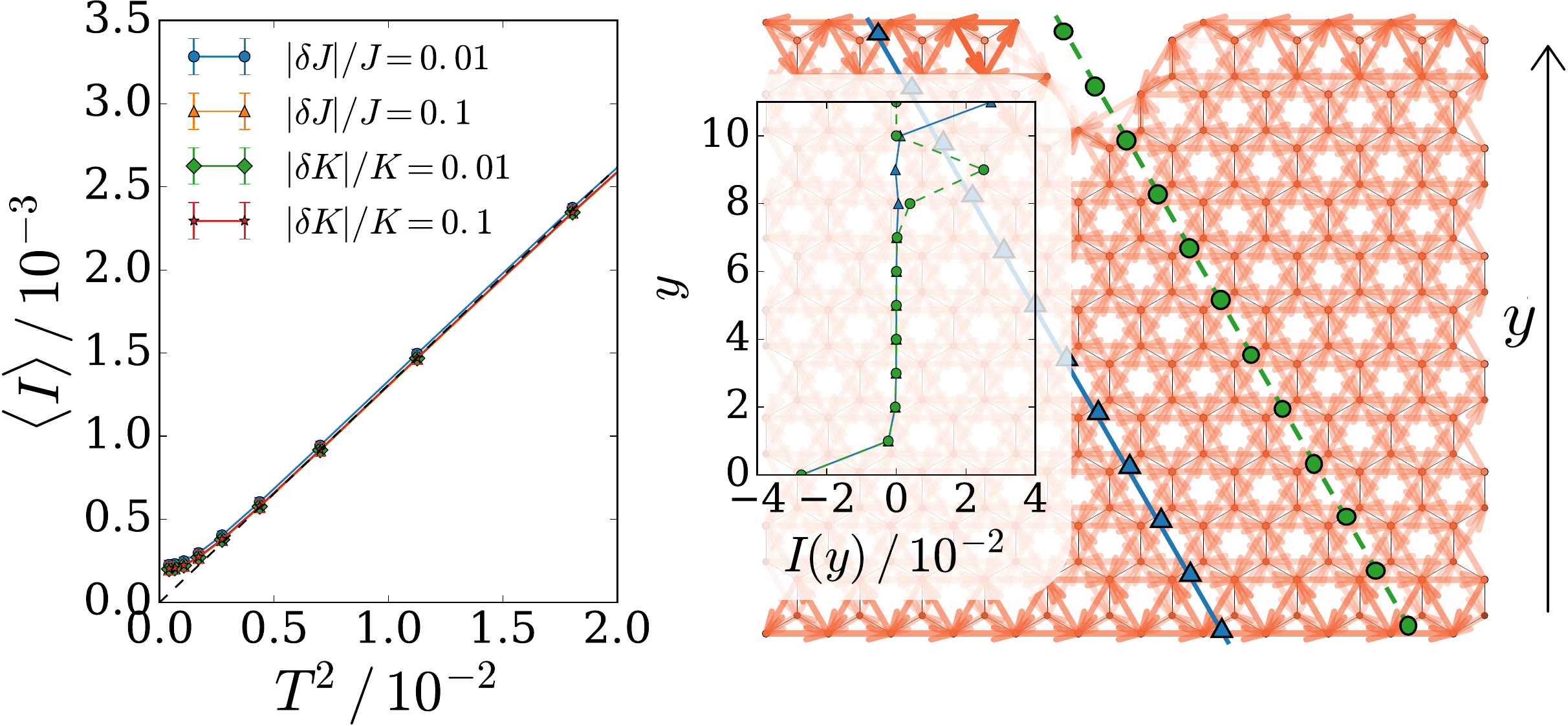}
\caption{Robustness of the edge currents to $J$ and $K$ disorder (Left) and lattice boundary defects (Right). In the (Left) plot we add a random component to either every $J$ term in the Hamiltonian or every $K$ term. We plot curves for $|\delta J|/J$ and $|\delta K|/K =$ 0.01 and 0.1, setting $J=1$ and $K=0.1$, for $L=40$. Each point is averaged over 100 disorder realizations. {The (Right) panel shows the effect of removing sites from the boundary. By plotting the current at the location of the defect (green circles) we see that the current shifts down onto the new edge. In both cases the total edge currents remain unaffected.}}
\label{fig:disorder}
\end{figure}

The origin of the energy current $I(T)$ is topological, so we expect it to be robust against local perturbations to the Hamiltonian. 
We have investigated this property in two settings. First we introduced disorder to the couplings $J$ and $K$.
For that we consider the no-vortex sector and add a random component to either every $J$ or $K$ value so that {\em e.g.} $J \rightarrow J + \delta J$ where $\delta J$ is a uniform random number between $-|\delta J|$ and $|\delta J|$. 
The currents are averaged over disorder $\langle I(T) \rangle$ and plotted as a function of $T^2$ for disorder strengths $|\delta J|/J$ and $|\delta K|/K$ of 1\% and 10\%, as shown in Fig.~\ref{fig:disorder} (Left). 
We see no impact of this disorder on the currents. 
Additionally, we introduce a boundary defect by removing sites from the edge. 
The energy currents near this defect are plotted in Fig.~\ref{fig:disorder} (Right). 
We find the impact of the defect is to divert the edge current around the missing lattice sites.
A current of equal intensity flows along the new edge. So the computation of the energy current gives the same $I(T)$ as without the defect.

\section{Zero temperature energy currents}
We now return to study the current at zero temperature $I(T=0)$ in more detail. 
As previously noted we find that at $T=0$ the current includes strong finite size effects as well as an offset $I_0$. 
We investigate these different elements by fitting values $I(T=0,L)$ computed for different system sizes $L$ to a scaling form
\begin{equation}
I(T=0,L) = I_0 + \frac{\Theta}{L^\gamma} \, .
\label{eqn:IT0-finite-size-scaling-form}
\end{equation}
Finite size scaling is also met in conformal field theory~\footnote{The energy current is predicted to be the $\tau-x$ component of the energy-momentum tensor\cite{DiF}. An elementary conformal mapping shows that on a cylinder $\langle T_{\tau x} \rangle=\pi c/ 12 L^2$.}.
At $T=0$ CFT predicts that the currents should depend on the circumference of the edge $L$ through the relation $I (L) = (\pi/12)\,c\,L^{-2}$. 

\begin{figure}[t]
		\includegraphics[width=\columnwidth]{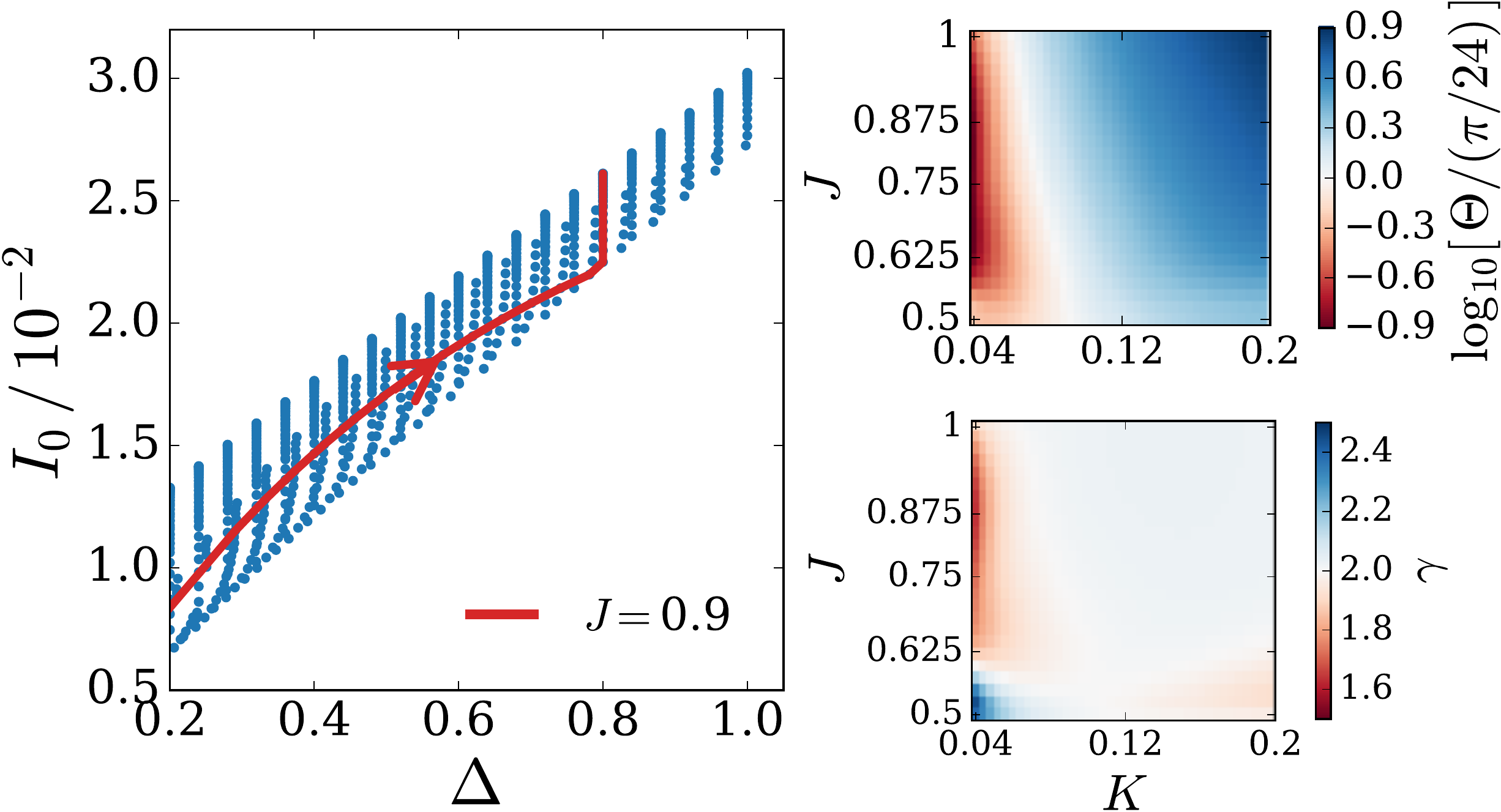}
	\caption{
	Finite size scaling analysis of the zero-temperature current in the no-vortex sector. This is carried out by fitting values $I(0, L)$ for $L = 10,12,\ldots,52$ to Eqn.~\eqref{eqn:IT0-finite-size-scaling-form}, while varying $J$ and $K$ within the no-vortex phase with $J_z = 1$. 
	(Left) We plot the extrapolated values $I_0$ as a function of the fermion gap $\Delta$. The relationship is non-unique. Onto the plot we have added the path of a single $J$ while $K$ increases.
	(Right) The pre-factor $\Theta$ and scaling exponent $\gamma$ are plotted over $J$ and $K$. We see for most parameters $\gamma \approx 2$. Interestingly, $\Theta$ is not found to have its CFT predicted value $\pi/12 \, c$ (where $c=1/2$) over most of the range investigated.%
	}
\label{fig:IT0-data}
\end{figure}

We plot the values obtained for $\gamma$, $\Theta$ and $I_0$ over a range of system parameters in the no-vortex phase in Fig.~\ref{fig:IT0-data}.
The fits are obtained by computing $I(T,L)$ at $T = 10^{-6}$ for a range of system sizes $L = 10,12,\ldots,52$ and fitting to Eqn.~\eqref{eqn:IT0-finite-size-scaling-form}. 
We vary $J$ and $K$ and find good agreement of $\gamma$ with the CFT scaling exponent $\gamma \approx 2$ for most $J$ and $K$. 
{The values of $\gamma$ noticably depart from 2 at small $K$. 
We attribute this to quasi-critical effects as the gap becomes small approaching the phase transition out of the gapped phase to the $K=0$ gapless phase}~\cite{Kitaev06,Kon2016}.
The extrapolated values $I_0$ show a complicated dependence on the system parameters.
In particular we note that by varying $K$ for constant $J$ we can alter the value of $I_0$ without changing the gap $\Delta$, as shown by the red trajectory plotted on Fig.~\ref{fig:IT0-data}  (Left).
In addition, the pre-factor $\Theta$ does not correspond to the CFT prediction $\Theta = (\pi/12)\,c$ over most of the range of $J$ and $K$. 
Hence, $I_0$ and $\Theta$ are strongly dependent on the microscopic parameters of the model, unlike the predictions of CFT. 
This demonstrates that, due to their chiral nature, the bulk states give a significant contribution to the energy current. Extracting the bulk contributions is crucial for the identification of the topological properties of the model.


\section{Conclusions} 

In conclusion we have presented a method to obtain the energy currents of a topological phase from a microscopic model. We have studied the topological properties of these currents at finite temperature and the corresponding transitions between different topological phases. Our results confirm predictions made using an effective description in terms of conformal field theory~\cite{Kitaev06}. Such CFT descriptions have been studied extensively at both the effective and microscopic level in the context of edge states of the quantum Hall effect~\cite{edge-QHE,Halperin1982}, but there is a major distinction between that setting and ours. Here, we have focused on the thermal transport appropriate to a topological superconductor as opposed to the charge currents found in the quantum Hall effect. 

We have seen that due to their CFT origin the energy currents are able to discriminate clearly between different topological phases at finite temperature. Moreover, we have demonstrated that these currents are very robust. Unlike the fragile physics of criticality that typically has zero measure in the parameter space, the central charges evaluated through edge energy currents are robust against significant variations of the Hamiltonian parameters. We demonstrated that, due to their topological origin, these currents are also robust against bulk disorder or even the introduction of boundary defects. This resilience makes them the ideal method to experimentally probe topological superconducting phases and reveal their conformal behavior~\cite{altimiras2009}.

\acknowledgments

C.N.S. would like to thank K. Meichanetzidis for helpful discussions. S.I. thanks M.A. Martin-Delgado for a fruitful conversation at an early stage of the present work. S.I. acknowledges funding form Spanish MINECO (grant MTM2014-54240-P), the QUITEMAD project/Comunidad de Madrid (S2013/ICE-2801). J.R.W. acknowledges funding from the NCCR QSIT. This project has received funding from the European Research Council (ERC) under the European Union's Horizon 2020 research and innovation programme (grant agreement No 648913). This work was supported in part by the EPSRC grant EP/I038683/1. This work was undertaken on ARC2, part of the High Performance Computing facilities at the University of Leeds, UK.

\appendix

\section{Defintion of the current operator}
Here we calculate an expression for the current operator first defined in Eqn.~\eqref{eqn-current-op},
\be
I_{jk} \equiv -i [h_k,h_j]. 
\nonumber
\ee
Let us write the two local energy terms as
\begin{eqnarray*}
h_k \, = \, \frac{i}{4} \sum_{l} A_{kl} c_k c_l & = & \frac{i}{4} \left( A_{kj} c_k c_j + \sum_{l \neq j} A_{kl} c_k c_l \right) \\
h_j \, = \, \frac{i}{4} \sum_{m} A_{jm} c_j c_m & = & \frac{i}{4} \left( A_{jk} c_j c_k + \sum_{m \neq k} A_{jm} c_j c_m \right) .
\end{eqnarray*}
Then expanding the commutator in this way we find
\begin{eqnarray*}
I_{jk} \! & = & \! \frac{i}{16} \left( A_{kj} A_{jk} \, [ c_k c_j, c_j c_k ] + \sum_{m \neq k} A_{kj} A_{jm} \, [ c_k c_j, c_j c_m ] \right. \\ 
& & \phantom{\bigg(} + \sum_{l \neq j} A_{kl} A_{jk} \, [ c_k c_l, c_j c_k ] + \sum_{l \neq j} A_{kl} A_{jl} \, [ c_k c_l, c_j c_l ] \\
& & \phantom{\bigg(} \left. + \sum_{l \neq j} \sum_{m \neq k,l} A_{kl} A_{jm} \, [ c_k c_l, c_j c_m ] \right) \, .
\end{eqnarray*}
The first term vanishes since $c_k c_j c_j c_k = c_j c_k c_k c_j = 1$, and similarly we can show that since the four indices in the last term are different that commutator also vanishes. Expanding the other terms yields
\begin{eqnarray}
I_{jk} & = & \frac{i}{8} \left( \sum_{m \neq k} A_{kj} A_{jm} \, c_k c_m + \sum_{l \neq j} A_{kl} A_{jk} \, c_l c_j  \right. \nonumber \\ 
& & \left. \qquad - \sum_{l \neq j} A_{kl} A_{jl} \, c_k c_j \right) \, .
\end{eqnarray}

We can rewrite this expression by exploiting the fact that for most $jk$ the elements $A_{jk}$ are zero. Define the {neighbourhood} $N(j)$ of a site $j$ as the set of sites $n_j$ where $A_{j \, n_j}$ is non-zero. Using this notation and looking back to our definition of $h_j$ and $h_k$ we can relabel $l \rightarrow n_k$ and $m \rightarrow n_j$. Additionally, we can notice that since the final term includes the product $A_{k \, n_k} A_{j \, n_k}$ that sum is restricted to sites $s$ in the overlap of $N(j)$ and $N(k)$. A final expression for the currents is then given by
\begin{eqnarray}
I_{jk} & = & -\frac{i}{8} \left( \sum_{n_k \neq j} A_{jk} A_{k \, n_k} \, c_j c_{n_k} - \sum_{n_j \neq k} A_{kj} A_{j \, n_j} \, c_k c_{n_j}  \right. \nonumber \\ 
& & \left. \qquad + \sum_{s \, \in N(j) \cap N(k)} A_{js} A_{sk} \, c_j c_k \right) \, .
\label{eqn-full-current-op}
\end{eqnarray}
This expression is more efficient to compute since the indices of the sums only run over a small set of sites rather than the whole lattice.

\section{Computing finite temperature expectation values}

We want to evaluate finite-temperature expectation values of the current operator given in Eqn.~\eqref{eqn-full-current-op}. The current is expressed in terms of two-point Majorana correlation functions $(i \, c_j c_k)$ and it is their expectation values we specifically need to compute. We can do this by going to the energy eigenbasis. 

The Hamiltonian of an arbitrary vortex sector of the Kitaev honeycomb is written as
\baleqn[b]
H &= \frac{i}{4} \, \sum_{j,k} \, A_{jk} \, c_j c_k \\
&= \sum_p \, \frac{\en_p}{2} \, \left( \, 2 \, f_p^\dagger f_p \,-\, 1 \, \right) \, ,
\ealeqn
where the first expression is in real-space and the second in the energy eigenbasis. The eigenvectors of the Hermitian matrix $(i A_{jk})$ encode the transformation from Majorana $c_j$ to fermionic diagonal modes $f_p$,
\begin{equation*}
c_i = \sum_p \,\left(\, [\underline{u}_p]_i \, f_p^\dagger + [\underline{u}_p]^*_i \, f_p \,\right)
\end{equation*}
where $\underline{u}_p$ is the eigenvector corresponding to the eigenvalue $-\en_p$ and the $(\ldots)^*$ in the second term is complex conjugation. Consider the two-point Majorana correlation functions, in the energy eigenbasis they are given by
\begin{align*}
i \, c_j c_k &= i\,\sum_{p,q} \,\left(\, [\underline{u}_p]_j \, f_p^\dagger + [\underline{u}_p]^*_j \, f_p \,\right) \,\left(\, [\underline{u}_q]_k \, f_q^\dagger + [\underline{u}_q]^*_k \, f_q \,\right) \\
&= i\, \sum_{p,q} \,\left(\, [\underline{u}_p]_j [\underline{u}_q]_k \, f_p^\dagger f_q^\dagger + [\underline{u}_p]_j [\underline{u}_q]^*_k \, f_p^\dagger f_q \right. \\ 
&\qquad\quad \left. + \, [\underline{u}_p]^*_j [\underline{u}_q]_k \, f_p f_q^\dagger + [\underline{u}_p]^*_j [\underline{u}_q]^*_k \, f_p f_q \, \right) \, \phantom{\sum_{p,q}} \\
&= i\, \sum_{p,q} \,\left(\, [\underline{u}_p \, (\underline{u}_q^*)^\dagger ]_{jk} \, f_p^\dagger f_q^\dagger + [\underline{u}_p (\underline{u}_q)^\dagger]_{jk} \, f_p^\dagger f_q \right.  \\
&\qquad\quad \left. + \, [\underline{u}_p (\underline{u}_q)^\dagger]^*_{jk} \, f_p f_q^\dagger + [\underline{u}_p (\underline{u}_q^*)^\dagger]^*_{jk} \, f_p f_q \, \right) \, . 
\end{align*}
We can easily compute the thermal expectation values in the energy eigenbasis. The only non-vanishing contributions are,
\baleqns
\langle f_p^\dagger f_q \rangle &= \frac{1}{\mathcal{Z}} \, \tr \! \left( \, e^{-\beta H} f_p^\dagger f_q \, \right) = \frac{1}{e^{\beta \en_p} + 1} \, \delta_{pq} \\
\langle f_p f_q^\dagger \rangle &= \frac{1}{\mathcal{Z}} \, \tr \! \left( \, e^{-\beta H} f_p f_q^\dagger \, \right) = \frac{e^{\beta \en_p}}{e^{\beta \en_p} + 1} \, \delta_{pq} \, .
\ealeqns
Further, we can use a property of the matrices in the equation for $(i \, c_j c_k)$, that $[\underline{u}_p (\underline{u}_p)^\dagger]^* = -[\underline{u}_p (\underline{u}_p)^\dagger]$ which must hold since $(i \, c_j c_k)$ is a hermitian operator, to obtain a final expression
\be
\langle \, i \, c_j c_k \, \rangle = - i \sum_p \, [\underline{u}_p (\underline{u}_p)^\dagger]_{jk} \, \tanh \left( \frac{\beta \en_p}{2} \right)
\ee

\bibstyle{plain}

\end{document}